\documentclass[12pt]{iopart}

\usepackage[dvips]{graphicx}
\usepackage{color}
\begin{document}
\newcommand{\E}{\mathrm{E}}
\newcommand{\Var}{\mathrm{Var}}
\newcommand{\bra}[1]{\langle #1|}
\newcommand{\ket}[1]{|#1\rangle}
\newcommand{\braket}[2]{\langle #1|#2 \rangle}
\newcommand{\be}{\begin{equation}}
\newcommand{\ee}{\end{equation}}
\newcommand{\ba}{\begin{eqnarray}}
\newcommand{\ea}{\end{eqnarray}}
\newcommand{\R}[1]{\textcolor{red}{#1}}
\newcommand{\B}[1]{\textcolor{blue}{#1}}
\title{Achieving ground state and enhancing entanglement by recovering information}

\author{Haixing Miao$^{1}$, Stefan Danilishin$^{2,3}$, Helge
M\"{u}ller-Ebhardt$^{3}$  and Yanbei Chen$^{4}$}
\address{$^{1}$School of Physics, University of Western Australia, WA 6009, Australia}
\address{$^{2}$Physics Faculty, Moscow State University, Moscow 119991, Russia}
\address{$^{3}$Max-Planck Institut f\"ur Gravitationsphysik (Albert-Einstein-Institut)
and Leibniz Universit\"at Hannover, Callinstr. 38, 30167 Hannover, Germany}
\address{$^{4}$Theoretical Astrophysics 130-33, California Institute of Technology,
Pasadena, CA 91125, USA}

\begin{abstract} For cavity-assisted optomechanical cooling experiments, it has been shown in the literature that
the cavity bandwidth needs to be smaller than the mechanical frequency in order to achieve the quantum
ground state of the mechanical oscillator, which is the so-called resolved-sideband or good-cavity limit.
We provide a new but physically equivalent insight into the origin of such a limit: that is information loss due to a finite 
cavity bandwidth. With an optimal feedback control to recover those information, we can surpass the resolved-sideband
limit and achieve the quantum ground state.  Interestingly, recovering those information can also
significantly enhance the optomechanical entanglement. Especially when the environmental temperature is high, the
entanglement will either exist or vanish critically depending on whether information is recovered
or not, which is a vivid example of a quantum eraser.
\end{abstract}

\maketitle

\section{Introduction}

Recently, achieving the quantum ground state of  a macroscopic mechanical oscillator has aroused great interests among
physicists. It will not only have significant impacts on quantum-limited measurements \cite{qmeas} but also will shed light
on quantum entanglements involving macroscopic mechanical degrees of freedom \cite{Bouwmeester,Mancini2,
Vitali, Paternostro, Helge, Hartmann}, which can be useful for future quantum computing and help us to understand
transitions between the classical and quantum domains \cite{Zurek, Diosi1, Penrose}.

By using a conventional cryogenic refrigeration, O'Connell {\it et al.} has successfully cooled a 6 GHz micromechanical
oscillator down to its ground state \cite{connell}. Meanwhile, to cool larger-size and lower-frequency mechanical oscillators at
high environmental temperature, there have been great efforts in trying different approaches: active feedback control and
parametrically coupling the oscillator to optical or electrical degrees of freedom \cite{Blair,Cohadon,Metzger,Naik, Gigan,Arcizet,Kleckner,Schliesser1, Corbitt1,Corbitt2,Schliesser2, Poggio,Favero,Teufel,Thompson,Lowry,Groblacher,
Schediwy,Jourdan, Aspelmeyer,LIGO, Schwab}.  The cooling mechanism has been extensively discussed, and certain
classical and quantum limits have been derived \cite{Vyatchanin_cooling, Mancini, Marquardt, Rae, Genes, ctrl, Diosi2, Rabl, zhao_prl}.
In the case of cavity-assisted cooling schemes, pioneering theoretical works by Marquardt {\it et al.} \cite{Marquardt} and
Wilson-Rae {\it et al.} \cite{Rae} showed that the quantum limit for the occupation number is $(\gamma/2\,\omega_m)^2$
\footnote{There is a factor of two difference in defining the cavity bandwidth here compared with the one
defined in Ref. \cite{Marquardt}.}. It dictates that, in order to achieve the ground state of the mechanical oscillator, the cavity
bandwidth  $\gamma$ must be smaller than the mechanical frequency $\omega_m$,  which is the so-called
resolved-sideband or good-cavity limit. This limit is derived by analyzing the quantum fluctuations of the radiation
pressure force on the mechanical oscillator. From a physically equivalent perspective, it can
actually be attributable to information loss: information of the oscillator motion leaks into the environment
without being carefully treated, which induces decoherence.

\begin{figure}[!h]
\includegraphics[width=0.5\textwidth, bb= 0 0 500 255, clip]{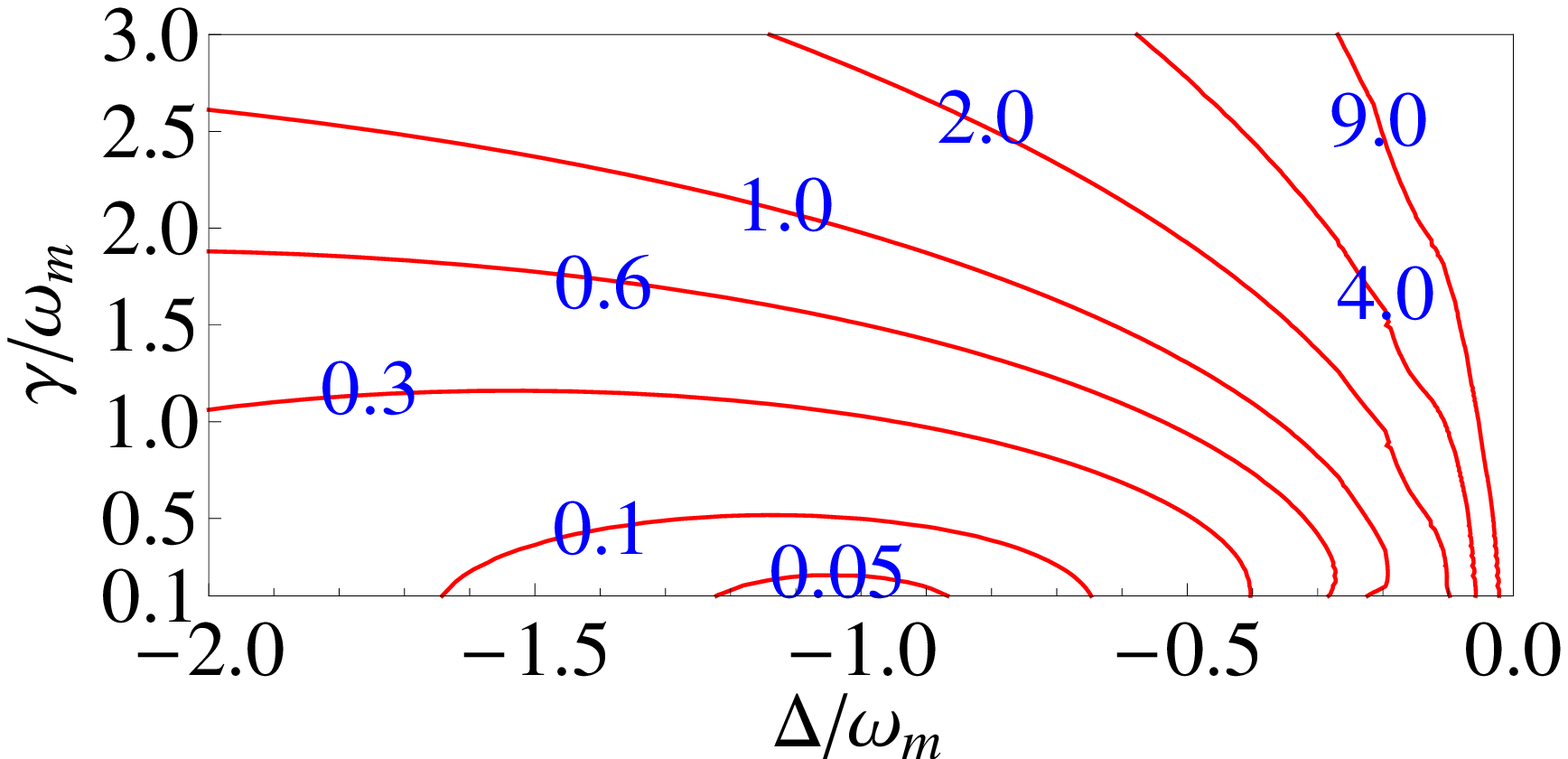}
\includegraphics[width=0.5\textwidth, bb= 0 0 500 255, clip]{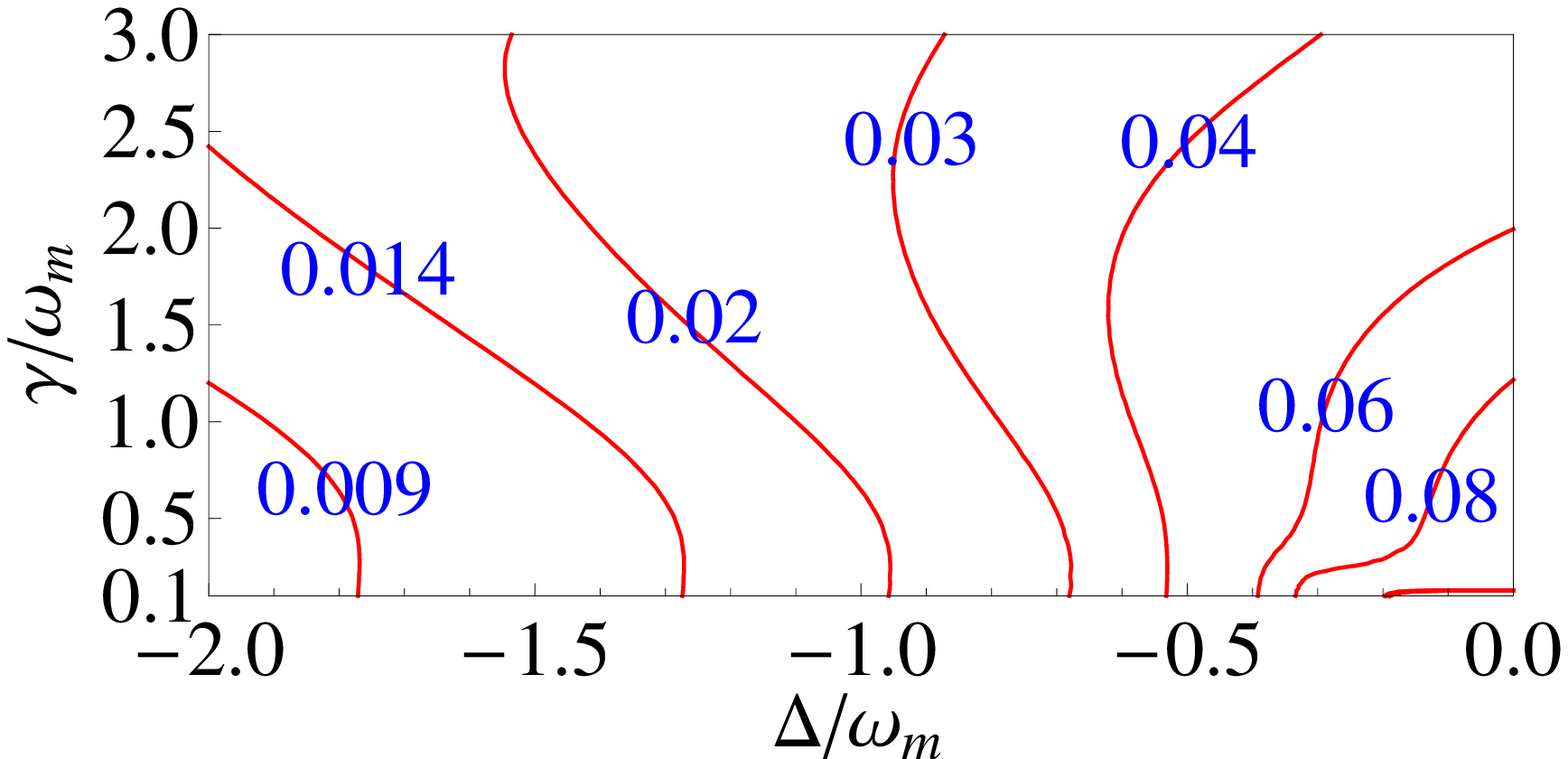}
\caption{A contour plot of the occupation number as a function of cavity bandwidth $\gamma$ and detuning $\Delta$
for the unconditional state (left)  as obtained in Ref. \cite{Marquardt, Rae} and optimally
controlled state (right) of which the details are in Sec. \ref{sec3}, \ref{sec4}, \ref{sec5}. \label{ctrl_state}}
\end{figure}

This perspective immediately illuminates two possible approaches for surpassing such a limit: (i) The
{\it first} one is to implement the novel scheme proposed by Elste {\it et al.} \cite{QN_int}, in which
the quantum noise gets destructively interfered and information of the oscillator motion around $\omega_m$
does not leak into the environment. Corbitt suggested an intuitive understanding by thinking of an optical
cavity with a movable front mirror rather than a movable end mirror in those cooling experiments \cite{Corbitt3}.
In this hypothetic scheme, optical fields directly reflected and those filtered through the cavity both contain
the information of the front-mirror motion. If the cavity detuning is appropriate,  these two bits of information
destructively interfere with each other, and the quantum coherence of the mechanical oscillator is maintained.
(ii) The {\it second} approach is to recover the information by detecting the cavity output. This will work
because a conditional quantum state---{\it best knowledge of the oscillator state conditional on the measurement
result}---is always pure for an ideal continuous measurement with no readout loss. Indeed, when the cavity bandwidth
is much larger than the mechanical frequency, the cavity mode will follow the oscillator dynamics and can therefore
be adiabatically eliminated. The quantum noise can be treated as being Markovian and a standard
stochastic-master-equation (SME) analysis has already shown that how the conditional quantum state approaches to
a pure state under a continuous measurement \cite{Hopkins, Gardiner, Milburn, Doherty1, Doherty2}. For a finite
cavity bandwidth considered here, the cavity mode has a dynamical timescale comparable to that of the mechanical
oscillator. Correspondingly, the quantum noise has correlations at different times and is non-Markovian. To estimate the
conditional state, a Wiener-filtering approach is more transparent than the SME \cite{state_pre}. As we will
show, the conditional quantum state of the oscillator in the cavity-assisted cooling schemes is indeed almost pure,  
with residue impurity contributed by the thermal noise, imperfections in detections and optomechanical entanglement 
between the oscillator and the cavity mode. In order to further localize the oscillator in the phase space and achieve 
its ground state,  an optimal feedback control is essential \cite{ctrl}. In Fig. \ref{ctrl_state}, the final occupation number 
of the unconditional state and optimally controlled state is shown. As long as the optimal control is applied, the minimally 
achievable occupation number of the oscillator will not be constrained by the resolved-sideband limit.

\begin{figure}
\includegraphics[width=0.65\textwidth, bb= -130 0 400 192, clip]{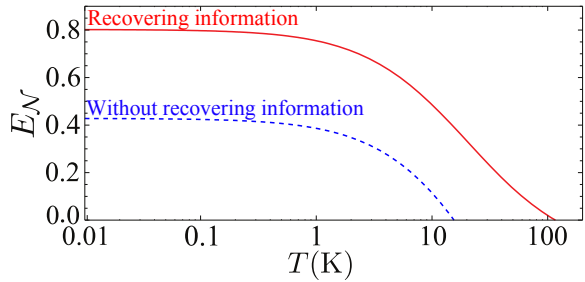}
\caption{Optomechanical entanglement strength $E_{\cal N}$ as a function of  temperature
$T$ with (solid) and without (dashed) recovering
information (details are in Sec. \ref{sec6}). \label{ENucT}}
\end{figure}

Another interesting issue in the optomechanical system is creating quantum entanglement between the cavity mode
and the oscillator, or even between two oscillators \cite{Mancini2, Bouwmeester, Vitali,  Paternostro, Helge, Hartmann}.
Intuitively, one might think that such an entanglement must be very vulnerable to the thermal decoherence, and the environmental
temperature needs to be extremely low in order to create it. However, as shown in Ref. \cite{Helge} and a more recent
investigation \cite{uni}, the environmental temperature---even though being an important factor---affects the entanglement
implicitly, and only the ratio between the interaction strength and thermal decoherence matters. The reason why in
Refs. \cite{Mancini2, Vitali, Paternostro, Hartmann}, the temperature plays a dominant role in determining the existence
of the entanglement can also be traced back to information loss, as briefly mentioned in Ref. \cite{uni}. Here we
will address this issue more explicitly.  Fig. \ref{ENucT} shows that by recovering the information contained in the
cavity output, the optomechanical entanglement can even be revived at high temperature. This is a vivid example
of a quantum eraser first proposed by Scully and Dr\"{u}hl \cite{q_eraser} and later demonstrated experimentally
\cite{q_eraser_exp}: Quantum coherence can be revived by recovering lost information.

The outline of this paper is the following: In Sec. \ref{sec2}, we will analyze the system dynamics by applying the
standard Langevin-equation approach and derive the spectral densities of important dynamical quantities. In Sec. \ref{sec3},
we obtain unconditional variances of the oscillator position and momentum, and evaluate the corresponding occupation
number, which recovers the resolved-sideband limit. In Sec. \ref{sec4}, conditional variances are derived via the 
Wiener-filtering approach, which clearly demonstrates that the conditional quantum state is almost pure. In Sec.
\ref{sec5}, we show the occupation number of the optimally controlled state and the corresponding optimal controller
to achieve it. In Sec. \ref{sec6}, we consider the optomechanical entanglement and demonstrate that significant
enhancements in the entanglement strength can be achieved after recovering information. In Sec. \ref{sec7}, to motivate
cavity-assisted cooling experiments, we consider imperfections in a real experiment and provide a numerical estimate for the
occupation number given a set of experimentally achievable specification. Finally, we conclude our main results in
Sec. \ref{sec8}.

\section{Dynamics and spectral densities\label{sec2}}

\begin{figure}[!h]
\includegraphics[width=0.45\textwidth, bb= -90 0 222 131, clip]{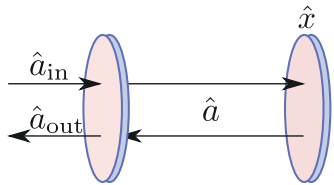}
\caption{A schematic plot of an optomechanical system with a mechanical oscillator $\hat x$ coupled to a cavity mode
$\hat a$ which in turn couples to the external ingoing $\hat a_{\rm in}$ and outgoing optical field $\hat a_{\rm out}$.
\label{fig1}}
\end{figure}
In this section, we will analyze the optomechanical dynamics and derive spectral densities of relevant quantities
which are essential for obtaining the occupation number of the mechanical oscillator.

\subsection{Dynamics}

Even though the dynamics of such a system has been discussed extensively in the literature \cite{Marquardt, Rae, Genes},
we will go through some equations for the coherence of this article. An optomechanical system and the relevant dynamical
quantities are shown schematically in Fig. \ref{fig1}. The corresponding Hamiltonian is given by
\be
\hat{\cal H}=\hbar\,\omega_c\,\hat a^{\dag}\hat a+\frac{\hat p^2}{2m}+\frac{1}{2}m\,\omega_m^2\hat x^2+\hbar\,G_0
\hat x\,\hat a^{\dag}\hat a+i\,\hbar\sqrt{2\gamma}\,(\hat a_{\rm in}e^{-i\,\omega_0\,t}\hat a^{\dag}-H.c.).
\ee
Here $\omega_c$ and $\omega_0$ are the cavity resonant frequency and the laser frequency, respectively; $\hat a$ is the
annihilation operator for the cavity mode, which satisfies $[\hat a,\,\hat a^{\dag}]=1$;
$\hat x$ and $\hat p$ denote the oscillator position and momentum with $[\hat x,\,\hat p]=i\,\hbar$; $m$ is mass of the oscillator;
$G_0\equiv \omega_0/L$ is the optomechanical coupling constant with $L$ the cavity length.
In the rotating frame at the laser frequency $\omega_0$,  a set of nonlinear Langevin equations can be obtained:
\begin{eqnarray}
\dot {\hat x}(t)&=\hat p(t)/m,\\
\dot {\hat p}(t)&=-\gamma_m\hat p(t)-m\,\omega_m^2\hat x(t)-\hbar\,G_0\hat a^{\dag}(t)\hat a(t)+\hat \xi_{\rm th}(t),\\
\dot{\hat a}(t)&=-(\gamma-i\Delta)\hat a(t)-i\,G_0 \hat x(t)\hat a(t)+\sqrt{2\gamma}\,\hat a_{\rm in}(t),
\end{eqnarray}
where cavity detuning $\Delta\equiv \omega_0-\omega_c$. To take into account the fluctuation-dissipation mechanism
of the oscillator coupled to a thermal heat bath at temperature $T$, we have included the mechanical damping $\gamma_m$
and the associated Brownian force $\hat \xi_{\rm th}$ of which the correlation function is $\langle{\hat \xi_{\rm th}(t)}\hat
\xi_{\rm th}(t')\rangle=2\,m\,\gamma_m k_B T\delta(t-t')$ in the high-temperature limit. In the cooling experiment, the 
cavity mode is driven by a coherent laser and, to a good approximation, the system is linear. To linearize the system, 
we simply replace any operator $\hat o(t)$ with the sum of a steady-state part and a small perturbed part, namely
$\hat o(t)\rightarrow\bar o+\hat o(t)$\footnote{For simplicity, we use the same $\hat o$ to denote its perturbed part.}.
We assume that the mean displacement of the oscillator is equal to zero with $\bar x=0$. The solution to $\bar a$ is simply
$\bar a=\sqrt{2\gamma}\,\bar a_{\rm in}/(\gamma-i\,\Delta)$ and $\bar a_{\rm in}=\sqrt{I_0/(\hbar\,\omega_0)}$ with $I_0$
the input optical power. We have chosen an appropriate phase reference such that
$\bar a$ is real and positive. The resulting linearized equations are
\begin{eqnarray}\label{5}
&m[\ddot{\hat x}(t)+\gamma_m \dot {\hat x}(t)+\omega_m^2\hat x(t)]=-\hbar\,\bar G_0[\hat a^{\dag}(t)+\hat a(t)]+
\hat \xi_{\rm th}(t),\\\label{6}
&\dot{\hat a}(t)+(\gamma-i\Delta)\hat a(t)=-i\,\bar G_0 \hat x(t)+\sqrt{2\gamma}\,\hat a_{\rm in}(t),
\end{eqnarray}
with $\bar G_0\equiv G_0\bar a$. The input-output relation of the cavity, which relates the cavity mode to the external continuum optical mode, reads \cite{Gardiner}
\be\label{7}
\hat a_{\rm out}(t)=\sqrt{\eta}[-\hat a_{\rm in}(t)+\sqrt{2\gamma}\,\hat a(t)]+\sqrt{1-\eta}\,\hat n(t),
\ee
where ${\eta}$ quantifies the quantum efficiency of the photodetector and $\hat n$ is the associated vacuum fluctuation
that is not correlated with $\hat a_{\rm in}$. The linearized dynamics of this system is fully described by Eqs. (\ref{5}), (\ref{6}) and (\ref{7})
which can be solved in the frequency domain. 

\noindent{\it Mechanical oscillator part.}---By denoting the Fourier component of any quantity $O$ as
$\tilde O(\Omega)$, the solution to the oscillator position is
\be\label{8}
\tilde x(\Omega)=\tilde R_{\rm eff}(\Omega)[\tilde F_{\rm BA}(\Omega)+\tilde \xi_{\rm th}(\Omega)].
\ee
Here the back-action force $\tilde F_{\rm BA}(\Omega)$ is
\be\label{9}
\tilde F_{\rm BA}(\Omega)=2\,\hbar\,\bar G_0\sqrt{\gamma}\,\chi(\Omega)[(\gamma-i\Omega)\tilde v_1(\Omega)-\Delta\,\tilde v_2(\Omega)],
\ee
where we have defined the amplitude quadrature $\tilde v_1(\Omega)$ and the phase quadrature $\tilde v_2(\Omega)$ of the 
vacuum fluctuation, namely $\tilde v_{1}(\Omega)\equiv[\tilde a_{\rm in}(\Omega)+\tilde a_{\rm in}^{\dag}(-\Omega)]/\sqrt{2}$
and $\tilde v_{2}(\Omega)\equiv[\tilde a_{\rm in}(\Omega)-\tilde a_{\rm in}^{\dag}(-\Omega)]/(i\sqrt{2})$.
Due to the well-known optical-spring effect, the mechanical response of the oscillator is modified from its original value $\tilde R_{xx}(\Omega)=-[m(\Omega^2+2\,i\,\gamma_m\Omega-\omega_m^2)]^{-1}$ to an effective one given by
\be\label{11}
\tilde R_{\rm eff}(\Omega)\equiv[\tilde R^{-1}_{xx}(\Omega)-\tilde \Gamma(\Omega)]^{-1}
\ee
with $ \tilde \Gamma(\Omega)\equiv2\,\hbar\,\bar G_0^2\,\Delta\,\chi$ and $\chi\equiv [(\Omega+\Delta+i\gamma)(\Omega-\Delta+i\gamma)]^{-1}$.

\noindent{\it Cavity mode part.}---The solution to the cavity mode is
\be
\tilde a(\Omega)=\frac{{\bar G_0\,\tilde x(\Omega)+i\sqrt{2\gamma}\,\tilde a_{\rm in}(\Omega)}}{{\Omega+\Delta+i\gamma}}.
\ee
In terms of amplitude and phase quadratures, it can be rewritten as
\ba
\tilde a_1(\Omega)&=\sqrt{2\gamma}\,\chi[(-\gamma+i\Omega)\tilde v_1(\Omega)+\Delta\,\tilde v_2(\Omega)]-\sqrt{2}\,\bar G_0\,\chi\,\Delta\, \tilde x(\Omega),\\
\tilde a_2(\Omega)&=\sqrt{2\gamma}\,\chi[-\Delta\,\tilde v_1(\Omega)-(\gamma-i\Omega)\tilde v_2(\Omega)]+\sqrt{2}\,\bar G_0 \,\chi \,(\gamma-i\Omega)\, \tilde x(\Omega).
\ea

{\it Cavity output part.}---Similarly, we introduce amplitude and phase quadratures for the cavity output: $\tilde Y_1(\Omega)\equiv[\tilde a_{\rm out}(\Omega)+\tilde a_{\rm out}^{\dag}(-\Omega)]/2$ and $\tilde Y_2(\Omega)\equiv[\tilde a_{\rm out}(\Omega)-\tilde a_{\rm out}^{\dag}(-\Omega)]/2$. Their solutions are
\be
\tilde Y_i(\Omega)=\tilde Y_i^{\rm vac}(\Omega)+\sqrt{\eta}\,\tilde R_{Y_iF}(\Omega)\,\tilde  x(\Omega), \quad(i=1,2).
\ee
The vacuum parts $\tilde Y_i^{\rm vac}$ of the output, which induce measurement shot noise, are the following:
\begin{eqnarray}\label{12}
\tilde Y_1^{\rm vac}(\Omega)&=\sqrt{1-\eta}\,\tilde n_1(\Omega)+\sqrt{\eta}\,\chi[(\Delta^2-\gamma^2-\Omega^2)\tilde v_1(\Omega)+2\,\gamma\,\Delta\,\tilde v_2(\Omega)],\\\label{13}
\tilde Y_2^{\rm vac}(\Omega)&=\sqrt{1-\eta}\,\tilde n_2(\Omega)+\sqrt{\eta}\,\chi[-2\,\gamma\,\Delta\,\tilde v_1(\Omega)+(\Delta^2-\gamma^2-\Omega^2)\tilde v_2(\Omega)].
\end{eqnarray}
The output response $\tilde R_{Y_i F}(\Omega)$ are defined as \cite{scaling_law}
\be
\tilde R_{Y_1 F}(\Omega)\equiv-2\sqrt{\gamma}\,\bar G_0\,\Delta\,\chi,\quad \tilde R_{Y_2 F}(\Omega)\equiv2\sqrt{\gamma}\bar G_0({\gamma-i\Omega})\chi.
\ee

\subsection{Spectral densities\label{2.2}}

Given the above solutions, we can analyze the statistical properties of the dynamical quantities.
We consider all noises to be Gaussian and stationary but not necessarily Markovian. Their statistical
properties are fully quantified by the spectral densities. We define a symmetrized single-sided spectral
density $\tilde S_{AB}(\Omega)$ according to the standard formula \cite{Kimble},
\be
2\pi\delta(\Omega-\Omega')\tilde S_{AB}(\Omega)=\langle A(\Omega)\tilde B^{\dag}(\Omega')\rangle_{\rm sym}=\langle \tilde A(\Omega)\tilde B^{\dag}(\Omega')+ \tilde B^{\dag}(\Omega')\tilde A(\Omega)\rangle.
\ee
For vacuum fluctuations $\hat a_{1,2}$, we simply have $\tilde S_{a_1a_1}(\Omega)=\tilde S_{a_2a_2}(\Omega)=1$ and $\tilde S_{a_1 a_2}(\Omega)=0$.

\noindent{\it Mechanical oscillator part.}---The spectral density for oscillator position is [cf. Eq.(\ref{8}) and Eq. (\ref{9})]
\be
\tilde S_{xx}(\Omega)=|\tilde R_{\rm eff}(\Omega)|^2 \tilde S_{FF}^{\rm tot}(\Omega),
\ee
with total force-noise spectrum
\be
\tilde S_{FF}(\Omega)=4\,\hbar\,m\,\Omega_q^3\,\gamma\,|\chi|^2(\gamma^2+\Omega^2+\Delta^2)+2\,\hbar\,m\,\Omega_F^2,
\ee
where we have introduced characteristic frequencies for the optomechanical interaction $\Omega_q\equiv (\hbar\,\bar G_0^2/m)^{1/3}$ and the thermal noise $\Omega_F\equiv \sqrt{2\gamma_m k_B T/\hbar}$. The spectral density for the oscillator momentum is simply $\tilde S_{pp}(\Omega)=m^2\,\Omega^2\tilde S_{xx}(\Omega)$.

\noindent{\it Cavity mode part.}---The spectral density for the cavity mode is a little bit complicated, which reads 
\be
{\bf S}_{aa}(\Omega)={\bf M}_0{\bf M}_0^{\dag}+{\bf M}_0{{\bf M}_1}^{\dag}+{\bf M}_1{{\bf M}_0}^{\dag}+{\bf M}_2 \tilde S_{xx}(\Omega).
\ee
Here the elements of the matrix ${\bf S}_{aa}$ are denoted by $\tilde S_{a_i a_j}(\Omega)\,(i,j=1,2)$; the matrix ${\bf M}_0$ is
\be
{\bf M}_{0}\equiv \sqrt{2\gamma}\,\chi \left[\begin{array}{cc}-\gamma+i\Omega&\Delta\\-\Delta&-\gamma+i\Omega\end{array}\right];
\ee
the matrix ${\bf M}_1$ is
\be
{\bf M}_1\equiv 2\sqrt{2}\hbar \bar G_0^2\sqrt{\gamma}|\chi|^2 \tilde R_{\rm eff}(\Omega)\left[\begin{array}{cc}-\Delta(\gamma-i\Omega)&\Delta^2\\(\gamma-i\Omega)^2&-\Delta(\gamma-i\Omega)\end{array}\right];
\ee
the matrix ${\bf M}_2$ is 
\be
{\bf M}_2\equiv 2\bar G_0^2|\chi|^2\left[\begin{array}{cc}\Delta^2&-\Delta(\gamma+i\Omega)\\-\Delta(\gamma-i\Omega)&\gamma^2+\Omega^2\end{array}\right].
\ee
The cross correlations between the cavity mode and the output $[{\bf S}_{aY}]_{ij}\equiv \tilde S_{a_i Y_j}(\Omega)$ are given by
\be
{\bf S}_{aY}={\bf M}_0{\bf M}_3^{\dag}+{\bf M}_0{\bf M}_1^{\dag}+{\bf M}_1{\bf M}_3^{\dag}+\sqrt{2\gamma}{\bf M}_2 \tilde S_{xx}(\Omega)
\ee
with
\be
{\bf M}_3\equiv\left[\begin{array}{cc}\Delta^2-\gamma^2-\Omega^2&2\gamma\Delta\\-2\gamma\Delta&\Delta^2-\gamma^2-\Omega^2\end{array}\right].
\ee
The cross correlation between the cavity mode and the oscillator is the following:
\be
\left[\begin{array}{c}\tilde S_{a_1 x}(\Omega)\\\tilde S_{a_2 x}(\Omega)\end{array}\right]=2\hbar \bar G_0\sqrt{\gamma}\chi^*\tilde R^*_{\rm eff}(\Omega)
{\bf M}_0\left[\begin{array}{c}\gamma+i\Omega\\-\Delta\end{array}\right]+\sqrt{2}\bar G_0 \chi
\left[\begin{array}{c}-\Delta\\\gamma-i\Omega\end{array}\right]\tilde S_{xx}(\Omega).
\ee
For the oscillator momentum, $\tilde S_{a_kp}(\Omega)=i\,m\,\Omega\,\tilde S_{a_k x} \;(k=1,2)$.

\noindent{\it Cavity output part.}---As an important feature of the quantum noise in this optomechanical system, there is a nonvanishing correlation between the shot noise $\hat Y_i^{\rm vac}$ and the
quantum back-action noise $\hat F_{\rm BA}$, and it has the following spectral densities [cf. Eqs. (\ref{9}), (\ref{12}) and (\ref{13})]:
\ba\label{18}
\tilde S_{FY_1^{\rm vac}}(\Omega)&=2\sqrt{\hbar\,m\,\gamma\,\eta\,\Omega_q^3}\,(\gamma+i\Omega)\,\chi^*,\\
\tilde S_{FY_2^{\rm vac}}(\Omega)&=2\sqrt{\hbar\,m\,\gamma\,\eta\,\Omega_q^3}\,\Delta\,\chi^*\label{19}
\ea
with $\chi^*$ the complex conjugate of $\chi$. Correspondingly, the spectral densities for the output quadratures read
\ba\nonumber
\tilde S_{Y_iY_j}(\Omega)&=\delta_{ij}+\eta\,\tilde R_{Y_iF}(\Omega)\tilde R_{xx}^{\rm eff}(\Omega)\tilde S_{FY_j^{\rm vac}}(\Omega)\\&+\eta\,[\tilde R_{Y_jF}(\Omega)\tilde R_{xx}^{\rm eff}(\Omega)\tilde S_{FY_i^{\rm vac}}(\Omega)]^*+\eta\,\tilde R_{Y_iF}(\Omega)\tilde R^*_{Y_jF}(\Omega)\tilde S_{xx}(\Omega).
\ea

The information of oscillator position $\hat x$ contained in the output $\hat Y_i$ are quantified by the
the cross correlations between $\hat x$ and $\hat Y_i$, which are
\be\label{21}
\tilde S_{x Y_i}(\Omega)=\sqrt{\eta}\,\tilde R_{xx}^{\rm eff}(\Omega)\tilde S_{FY_i^{\rm vac}}(\Omega)+\sqrt{\eta}\,\tilde R_{Y_i F}^*(\Omega)\tilde S_{xx}(\Omega).
\ee
Similarly, for the oscillator momentum, $\tilde S_{pY_k}(\Omega)=-i\,m\,\Omega\,\tilde S_{x Y_k}(\Omega)\;(k=1,2)$.

\section{Unconditional quantum state and resolved-sideband limit\label{sec3}}

\begin{figure}[!h]
\includegraphics[width=0.8\textwidth, bb= -116 0 464 88, clip]{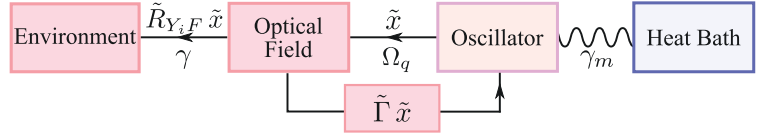}
\caption{A block diagram for the optomechanical system. The optomechanical cooling
can be viewed as a feedback mechanism and reduce the thermal occupation number of the oscillator which has
an effective temperature much lower than the heat bath. Meanwhile, some information of the oscillator motion
flows into the environment without being properly treated, leading to the resolved-sideband limit. \label{info_flow}}
\end{figure}
In the red-detuned regime ($\Delta<0$) where
those cavity-assisted cooling experiments are currently working, a delayed response of the cavity
mode to the oscillator motion gives rise to a viscous damping which can significantly reduce
the thermal occupation number of the oscillator, as shown schematically in Fig. \ref{info_flow}.
Physically, it is because the mechanical response is changed into an effective one [cf. Eq. (\ref{11})] 
while the thermal force spectrum remains the same. The ground state can be achieved when the
occupation number is much smaller than one. If we neglect the information of the oscillator motion
that contains in the output, the resulting quantum state of the oscillator will be {\it unconditional} and the corresponding 
occupation number of the oscillator can be obtained with the following standard definition:
\be\label{23}
{\cal N}\equiv\frac{1}{\hbar\,\omega_m}\left(\frac{V_{pp}}{2\,m}+\frac{1}{2}m\,\omega_m^2V_{xx}\right)-\frac{1}{2},
\ee
where variances of the oscillator position $V_{xx}$ and momentum $V_{pp}$ are related to the spectral densities
by the following formula:
\be
V_{xx,pp}=\int_{0}^{\infty}\frac{d\Omega}{2\pi}\,\tilde S_{xx,pp}(\Omega).
\ee
Since ${\cal N}$ is dimensionless, it only depends on the following ratios
\be
\Omega_q/\omega_m,\,\gamma/\omega_m,\, \Delta/\omega_m,\,\Omega_F/\omega_m, \gamma_m/\omega_m.
\ee
The oscillator mass and frequency only enter implicitly. As long as those ratios are
the same in different experiments, the final achievable thermal occupation number of different oscillators will
be identical.

The resulting $\cal N$ is shown in the left panel of Fig. \ref{ctrl_state}. To highlight the quantum limit, we have
fixed the interaction strength $\Omega_q$ with $\Omega_q/\omega_m=0.5$, and we have neglected the thermal 
force noise. In the optimal cooling regime with $\Delta=-\omega_m$, a simple closed form for the occupation 
number can be obtained \cite{Genes}
\be
{\cal N}=\gamma^2/(2\,\omega_m)^2+\frac{[1+(\gamma/\omega_m)^2]
(\Omega_q/\omega_m)^3}{4[1+(\gamma/\omega_m)^2-2(\Omega_q/\omega_m)^3)]}.
\ee
The resolved-sideband limit is achieved for a weak interaction strength $\Omega_q\rightarrow 0$ and
\be
{\cal N}_{\rm lim}=\gamma^2/(2\,\omega_m)^2.
\ee
In the next section, we will demonstrate that such a limit can indeed be surpassed by recovering the information
contained in the cavity output.

\section{Conditional quantum state and Wiener filtering\label{sec4}}

Since, given a a finite cavity bandwidth, the cavity output contains the information of the oscillator position [cf. Eq. (\ref{21})], 
according to the quantum mechanics, measurements of the output will collapse the oscillator wave function and project it into 
a {\it conditional} quantum state that is in accord with the measurement result. The conditional state or equivalently its Wigner
 function is completely determined by the conditional mean $[x^{\rm cond},\, p^{\rm cond}]$ and the covariance matrix
${\bf V}^{\rm cond}$ between the position and momentum. More explicitly, the Wigner function reads
\be
W(x,p)=\frac{1}{2\pi\sqrt{\det {\bf V}^{\rm cond}}}\exp\left[-\frac{1}{2}\delta \vec X\, {{\bf V}^{\rm cond}}^{-1}\delta \vec X^{T}\right]
\ee
with $\delta \vec X=[x-x^{\rm cond},\,p-p^{\rm cond}]$.
Since more information is acquired, the conditional quantum state is always
more pure than the unconditional counterpart. In the limiting case of an ideal measurement,
the conditional quantum state of the mechanical
oscillator would be pure with variances constrained by the {\it Heisenberg Uncertainty}, i.e., $\det {\bf V}^{\rm cond}|_{\rm pure\,\,state}=\hbar^2/4$.

To derive the conditional mean and variances, a usually applied mathematical tool is the stochastic-master-equation (SME),
which is most convenient for treating Markovian process \cite{Hopkins, Gardiner, Milburn, Doherty1, Doherty2}. In the case
considered here, however, the cavity has a bandwidth comparable to the mechanical frequency, and the quantum noise
is non-Markovian. The corresponding conditional mean and variance can be derived more easily with
the Wiener-filtering approach. As shown in Ref. \cite{state_pre}, the conditional mean of any quantity $\hat o(t)$ given
certain measurement result $Y(t')\,(t<t')$ can be written as
\be
o(t)^{\rm cond}\equiv \langle\hat o(t)\rangle^{\rm cond}=\int_{-\infty}^tdt'\, K_{o}(t-t')Y(t').
\ee
Here $K_o(t)$ is the optimal Wiener filter and is derived by using the standard Wiener-Hopf method. Its frequency representation is
\be\label{27}
\tilde K_o(\Omega)=\frac{1}{\tilde \psi_+(\Omega)}\left[\frac{\tilde S_{oY}(\Omega)}{\tilde \psi_-(\Omega)}\right]_+
\equiv \frac{\tilde G_o(\Omega)}{\tilde\psi_+(\Omega)},
\ee
where $[\,]_{+}$ means taking the causal component and $\tilde \psi_{\pm}$ is a spectral factorization
of the output $\tilde S_{YY}\equiv\tilde \psi_+\tilde \psi_-$ with $\tilde \psi_{+}$ ($\tilde \psi_{+}$)
and its inverse analytical in the upper-half (lower-half) complex plane and we have introduced $\tilde G_o(\Omega)$.
The conditional covariance between $\hat A$ and $\hat B$ is given by
\begin{eqnarray}\nonumber
V^{\rm cond}_{A B}&\equiv& \langle \hat A(0) \hat B(0) \rangle_{\rm sym}^{\rm cond}-\langle \hat A(0)\rangle^{\rm cond} \langle \hat B(0)\rangle^{\rm cond} \\&=&\int_0^{\infty}\frac{d\Omega}{2\pi}\left[\tilde S_{AB}(\Omega)-\tilde G_{A}(\Omega)\tilde G^*_{B}(\Omega)\right]\label{28}.
\end{eqnarray}
Since the first term is the unconditional variance, the second term can be interpreted as reductions in
the uncertainty due to acquiring additional information from the measurement.

Those results can be directly applied to the optomechanical system. Suppose we
measure the following quadrature of the cavity output
\be
\hat Y_{\zeta}=\hat Y_1\sin \zeta+\hat Y_2\cos\zeta
\ee
and its spectral density is
\be
\tilde S_{YY}(\Omega)=\tilde S_{Y_1Y_1}(\Omega)\sin^2\zeta+\Re[\tilde S_{Y_1Y_2}(\Omega)]\sin(2\zeta)+\tilde S_{Y_2Y_2}(\Omega)\cos^2\zeta
\ee
The cross correlation between $\hat Y_{\zeta}$ and oscillator position (momentum) is simply
\be
\tilde S_{xY,pY}= \tilde S_{x Y_1, pY_1}(\Omega)\sin\zeta+\tilde S_{x Y_2, p Y_2}(\Omega)\cos\zeta.
\ee
Plugging the spectral densities $\tilde S_{Y_iY_j}, \,\tilde S_{xY_i,p Y_i}$ and $\tilde S_{xx,pp}$
derived in subsection \ref{2.2} into Eq. (\ref{28}), we can obtain the conditional
covariances of the oscillator position and momentum, namely $V_{xx}^{\rm cond}$, $V_{pp}^{\rm cond}$ and $V_{xp}^{\rm cond}$.
\begin{figure}
\includegraphics[width=0.65\textwidth, bb= -160 0 500 255, clip]{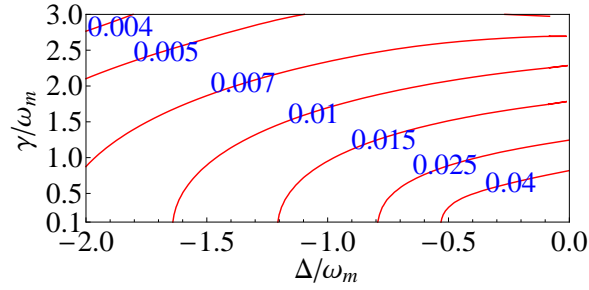}
\caption{A contour plot for the effective occupation number of the conditional quantum state.
For comparison, we have chosen the same specification as in the unconditional case.\label{cond}}
\end{figure}

To quantify how pure the conditional quantum state is, the occupation number defined in
Eq. (\ref{23}) is no longer an adequate summarizing figure. This is because generally
$V_{xp}^{\rm cond}$ is not equal to zero, and a pure squeezed state can have a large 
occupation number defined in Eq. (\ref{23}). A well-defined figure of merit is the 
{\it uncertainty product}, which is given by
\be
U\equiv \frac{2}{\hbar}\sqrt{V_{xx}^{\rm cond}V_{pp}^{\rm cond}-{V_{xp}^{\rm cond}}^2}.
\ee
From it, we can introduce an effective occupation number
\be
{\cal N}_{\rm eff}=(U-1)/2,
\ee
which quantifies how far the quantum state deviates from the pure one with ${\cal N}_{\rm eff}=0$. It is
identical to the previous definition [cf. Eq. (\ref{23})] in the limiting case
of $V^{\rm cond}_{xx}=V^{\rm cond}_{pp}/(m^2\omega_m^2)$ and $V_{xp}^{\rm cond}=0$, which is
actually satisfied in most of the parameter regimes plotted in Fig. \ref{ctrl_state}.

For a numerical estimate and comparing with the unconditional quantum state in the previous section,
we assume the same specification and an ideal phase quadrature detection with $\zeta=0$ and $\eta=1$.
The resulting effective occupation number is shown in Fig. \ref{cond}. Just as expected, the conditional
quantum state is not constrained by the resolved-sideband limit and is almost independent of detailed
specifications of $\gamma$ and $\Delta$. The residue occupation number or impurity of the state, shown in Fig. \ref{cond}, is due
to information of the oscillator motion being confined inside the cavity. Such a confinement
is  actually attributable to the quantum entanglement between the cavity mode and the oscillator, as we will discuss in
Sec. \ref{sec6}.

\section{Optimal feedback control\label{sec5}}

\begin{figure}[!h]
\includegraphics[width=0.8\textwidth, bb= -114 0 459 125, clip]{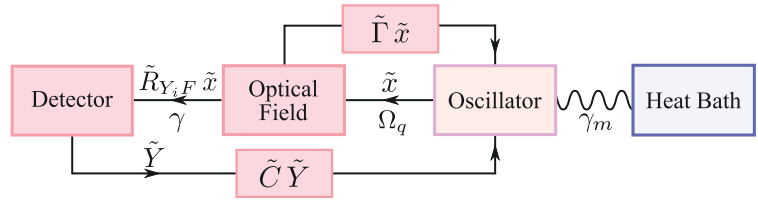}
\caption{A block diagram for the feedback control scheme.
A force is applied onto the mechanical oscillator based on the measurement result with a control
kernel $\tilde C$. In the detuned case ($\Delta\neq0$), the radiation pressure and the control force
work together to place the mechanical oscillator near its quantum ground state.  \label{control}}
\end{figure}

Even though the conditional quantum state has minimum variances in position and momentum, the oscillator itself
actually wanders around in phase space with its center given by the conditional mean $[x^{\rm cond}(t),\,p^{\rm cond}(t)]$ at any instant $t$.
In order to localize the mechanical oscillator and achieve its ground state, we need to apply a feedback control, i.e., a
force onto the oscillator, according to the measurement result. Such a procedure is shown schematically in Fig. \ref{control}.
Depending on different controllers, the resulting controlled state will have different occupation numbers.
The minimum occupation number can only be achieved if the unique optimal controller is applied. In Ref. \cite{ctrl},
the optimal controller was derived for a general linear continuous measurement. It can be directly applied
to the optomechanical system with non-Markovian quantum noise considered here.

Specifically, given the measured output quadrature $\hat Y_{\zeta}$,
the feedback force applied to the oscillator can be written in the time and the frequency domains as:
\be
\hat F_{\rm FB}(t) = \int_{-\infty}^tdt'\,C(t-t')\hat Y_\zeta(t')\,,\mbox{ and }\tilde F_{\rm FB}(\Omega) = \tilde C(\Omega)\tilde Y_\zeta(\Omega)\,.
\ee
with $C(t)$ a causal control kernel. 
The equation of motion for the oscillator will be modified as [cf. Eq. (\ref{6})]
\be
m[\ddot{\hat x}_{\rm ctrl}(t)+\gamma_m \dot {\hat x}_{\rm ctrl}(t)+\omega_m^2\hat x_{\rm ctrl}(t)]=-\hbar\,\bar G_0[\hat a^{\dag}(t)+\hat a(t)]+\hat \xi_{\rm th}(t)+\hat F_{\rm FB}(t).
\ee
In the frequency domain, the controlled oscillator position $\hat x_{\rm ctrl}$ is related to the uncontrolled one $\hat x$ by 
\be\label{Kdef}
\tilde x_{\rm ctrl}(\Omega)=\tilde x(\Omega)+\frac{\tilde R^{\rm eff}_{xx}(\Omega)\tilde C(\Omega)\tilde Y_{\zeta}(\Omega)}{1-\tilde R^{\rm eff}_{xx}(\Omega)\tilde R_{Y_\zeta F}(\Omega)\tilde C(\Omega)}.
\ee
As shown in Ref. \cite{ctrl}, by minimizing the effective occupation number of the controlled state,
the optimal controller can be derived and it is given by
\be
\tilde C^{\rm opt}(\Omega)=-\frac{{\tilde R_{xx}^{\rm eff}(\Omega)}^{-1}\tilde K^{\rm opt}_{\rm ctrl}(\Omega)}{1-\tilde R_{YF}(\Omega)\tilde K^{\rm opt}_{\rm ctrl}(\Omega)},
\ee
where
\be
\tilde K^{\rm opt}_{\rm ctrl}(\Omega)=\frac{1}{\tilde \psi_+(\Omega)}\left[\tilde G_x(\Omega)-\frac{G_x(0)}{\sqrt{V_{pp}^{\rm cond}/V_{xx}^{\rm cond}}-i\,\Omega}\right]
\ee
with $\tilde G_x(\Omega)=[{\tilde S_{xY}(\Omega)}/{\tilde \psi_{-}(\Omega)}]_+$ as defined in Eq. (\ref{27}).

From Eq. (\ref{Kdef}), we can find out the spectral densities and the covariance for the controlled position and momentum.
As it turns out, there is an intimate connection between the optimally controlled state and the conditional quantum state.
Due to the requirement of stationarity, it indicates that $V_{xp}^{\rm ctrl}=0$ [$V_{xp}=(1/2)m \dot V_{xx}(0)=0$], and therefore
the optimally controlled state is always less pure than the conditional state.
The corresponding purity of the optimally controlled state reads  \cite{ctrl}
\be
\label{opt}
U_{\rm ctrl}^{\rm opt}= \frac{2}{\hbar}\sqrt{V_{xx}^{\rm ctrl}V_{pp}^{\rm ctrl}}|_{\rm optimally\;controlled}=
\frac{2}{\hbar}\left[\sqrt{V_{xx}^{\rm cond} V_{pp}^{\rm cond}}+ |V_{xp}^{\rm cond}|\right]\,.
\ee
The occupation number ${\cal N}$  for the optimally controlled state was shown in Fig. \ref{ctrl_state} in the introduction. 
Since $V_{xp}^{\rm cond}$ is quite small compared with $V_{xx,pp}^{\rm cond}$, the resulting occupation number is very close to that of the conditional quantum state. Therefore, as long as the optimal controller is applied, the mechanical oscillator is almost in its quantum ground state and the resolved-sideband limit does not impose significant constraints.

\section{Conditional Optomechanical Entanglement and Quantum Eraser\label{sec6}}

In this section, we will analyze the optomechanical entanglement between the oscillator and the cavity mode. In particular, we will show
(i) the residue impurity of the conditional quantum state of the oscillator is induced by this optomechanical entanglement; (ii)
if the environmental temperature is high, the existence of entanglement critically depends on whether the information in the cavity output is recovered or not. In other words, the quantum correlation is affected by the ``eraser" of certain information, which manifests the idea of ``quantum eraser" proposed by Scully and Dr\"{u}hl \cite{q_eraser}.

\begin{figure}
\includegraphics[width=0.5\textwidth, bb=0 0 500 255, clip]{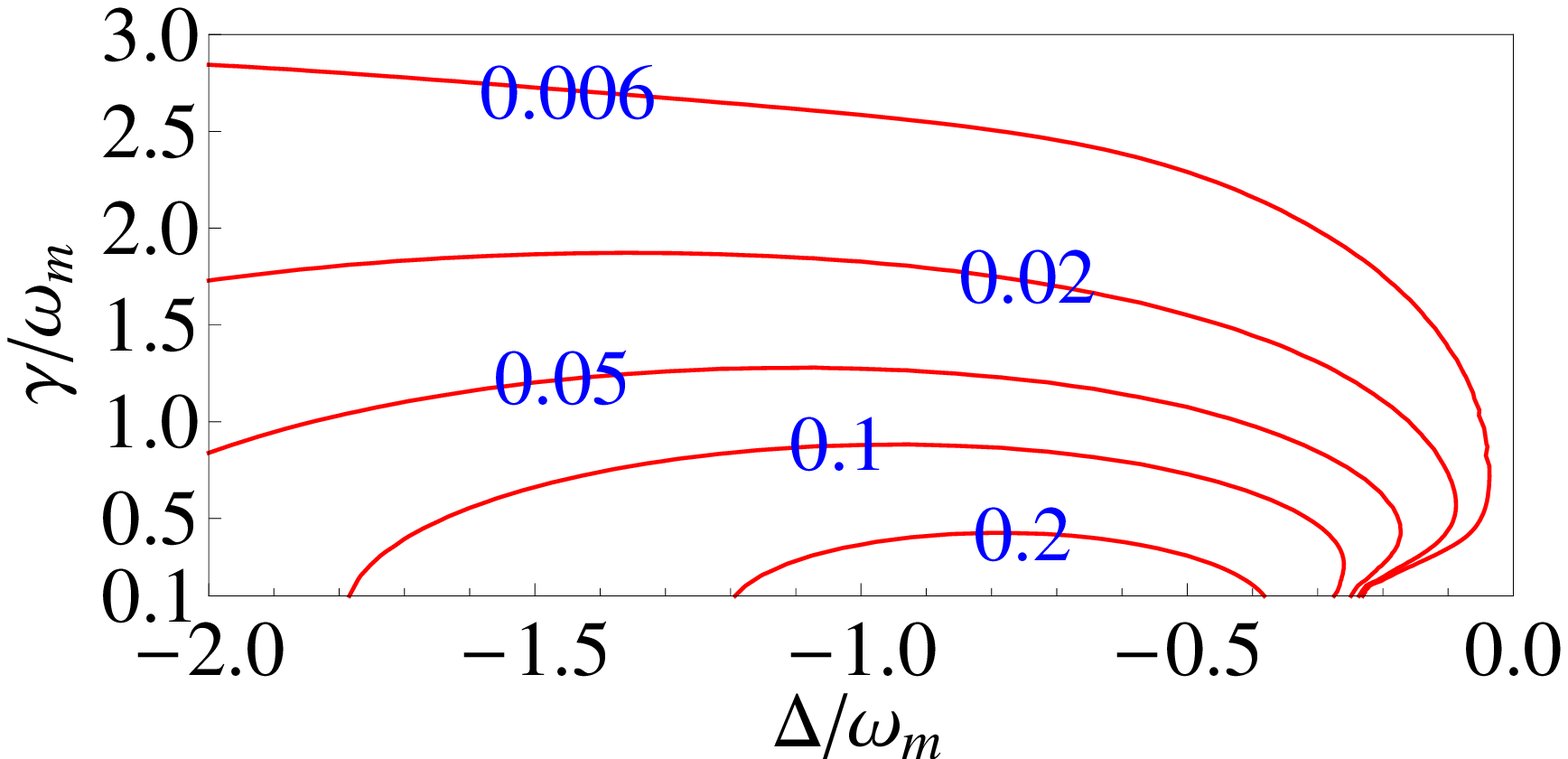}
\includegraphics[width=0.5\textwidth, bb=0 0 500 255, clip]{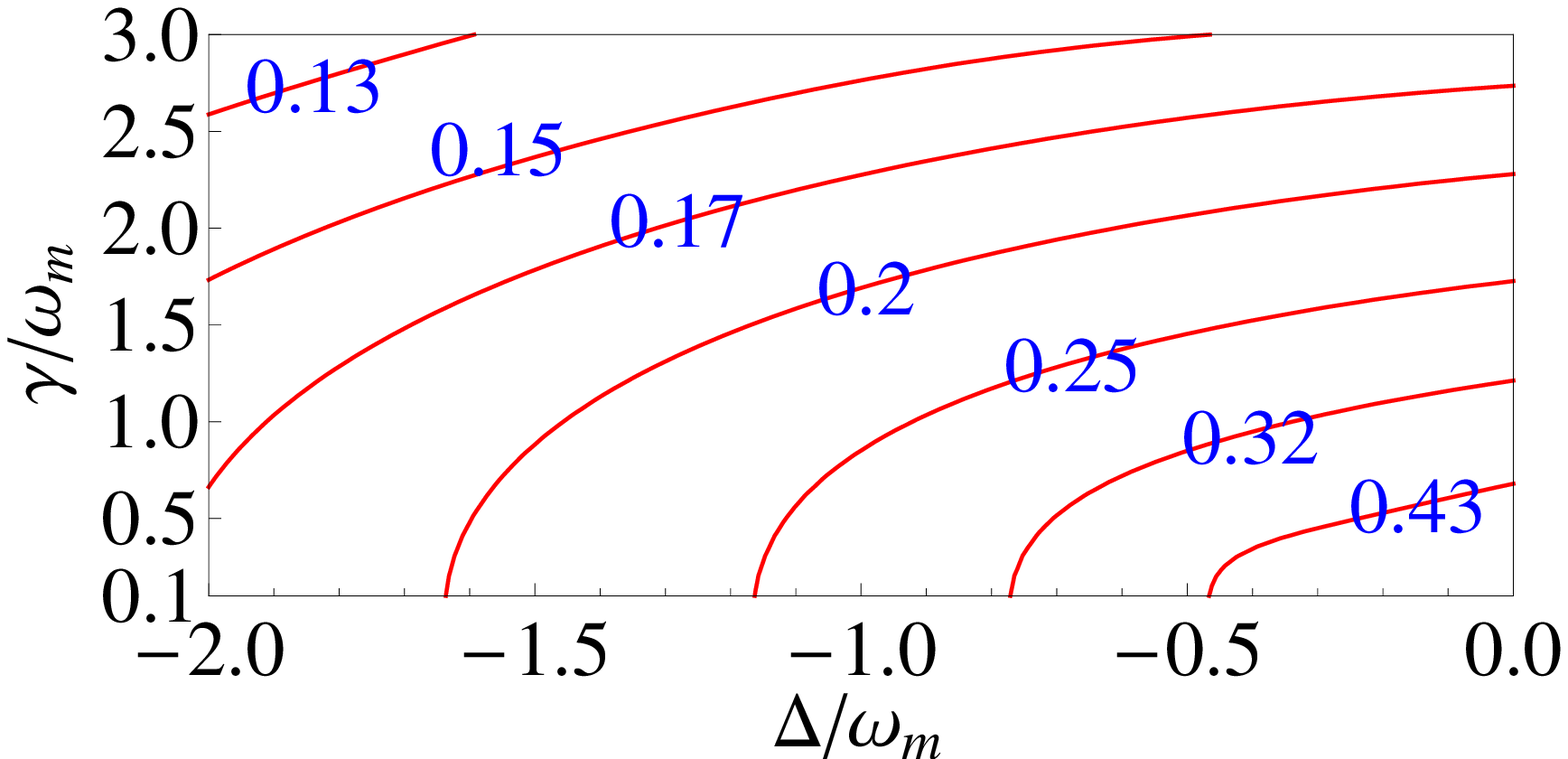}
\caption{Contour plots of the logarithmic negativity $E_{\cal N}$ for unconditional (left) and conditional (right) entanglement
between the cavity mode and the oscillator. We have assumed that $\Omega_q/\omega_m=0.5$ to make sure that the resulting
optomechanical system is stable in those parameter regimes shown in the figure. Besides, to manifest the entanglement,
we have ignored thermal noise. \label{ENu}}
\end{figure}
The existence of optomechanical entanglement is shown in the pioneering work by Vitali {\it et al.} \cite{Vitali}. The entanglement criterion, i.e., inseperability, is based upon positivity of the partially transposed density matrix \cite{Peres,Horodecki,Simon}. In the case of Gaussian variables considered here, this reduces to the following uncertainty principle in phase space:
\be
{\bf V}_{\rm pt}+\frac{1}{2}{\bf  K}\ge0,\quad {\bf K}=\left(\begin{array}{cc}0&-2i\\2i&0\end{array}\right)
\ee
with ${\bf K}$ denoting the commutator matrix. Partial transpose is equivalent to time reversal and the momentum of the oscillator changes sign. The corresponding partially transposed covariance matrix ${\bf V}_{\rm pt}={\bf V}|_{\hat p\rightarrow -p}$. From the Williamson theorem, there exists a symplectic transformation ${\bf S}\in S_{p(4,\bf{R})}$
such that ${\bf S}^{\rm T}{\bf V}_{\rm pt}{\bf S}=\bigoplus^{2}_{i=1}{{\rm Diag}[\lambda_i,\,\lambda_i]}$.
Using the fact that ${\bf S}^{\rm T}{\bf K}{\bf S}={\bf K}$, the above uncertainty principle requires
$\lambda_i\ge 1$. If $\exists\lambda< 1$, the states are entangled.
The amount of entanglement can be quantified by the logarithmic negativity $E_{\cal N}$ \cite{Vidal, Adesso}, which is defined as
\be\label{ENdef}
E_{\cal N}\equiv\max[-\ln\lambda,\,0].
\ee
Given a $4\times 4$ covariance matrix $\bf V$ between the oscillator $[\hat x,\,\hat p]$ and the cavity mode $[\hat a_1,\,\hat a_2]$, the simplectic eigenvalue $\lambda$ has the following closed form:
\be
\lambda= \sqrt{\Sigma-\sqrt{\Sigma^2-4\det {\bf V}}}/\sqrt{2},
\ee
where $\Sigma\equiv \det {\bf A}+\det {\bf B}-2\det {\bf C}$ and
\be
{\bf V}=\langle [\hat x,\hat p,\hat a_1,\hat a_2]^{T}[\hat x,\hat p,\hat a_1,\hat a_2]\rangle_{\rm sym}=\left[\begin{array}{cc}{\bf A}_{2\times 2}&{\bf C}_{2\times 2}\\
{\bf C}_{2\times 2}^{\bf T}&{\bf B}_{2\times 2}\end{array}\right].
\ee

In Ref. \cite{Vitali}, the information contained in the cavity output was ignored and unconditional
covariances were used to evaluate the entanglement measure $E_{\cal N}$. We can call it unconditional entanglement. If the information were recovered, conditional covariances obtained in Eq. (\ref{28}) will replace the unconditional counterparts. In Fig. \ref{ENu}, we compare
the unconditional and conditional entanglement, and it clearly shows that the entanglement strength increases dramatically in the conditional case.  Additionally, the regime where the entanglement is strong is in accord with where the conditional quantum state of the oscillator is less pure as shown in Fig. \ref{cond}. Indeed, there is a simple analytical relation between the effective occupation number ${\cal N}_{\rm eff}$
and the logarithmic negativity $E_{\cal N}$ in this ideal case with no thermal noise---that is
\be
E_{\cal N}=-2\ln\left[\sqrt{{\cal N}_{\rm eff}+1}-\sqrt{{\cal N}_{\rm eff}}\right]\approx 2\sqrt{{\cal N}_{\rm eff}}
\ee
for small ${\cal N}_{\rm eff}$  \cite{Vidal}.  Therefore, the limitation of a cooling experiment actually comes from the optomechanical entanglement, which justifies our claim in Sec. \ref{sec4}.

\begin{figure}
\includegraphics[width=0.65\textwidth, bb= -130 0 405 200, clip]{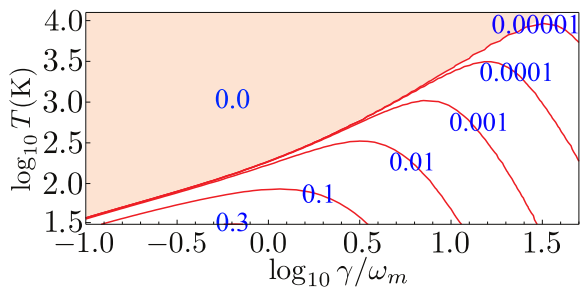}
\caption{Logarithmic negativity $E_{\cal N}$ as a function of cavity bandwidth and environmental temperature.
We have chosen $\Omega_q/\omega_m=1$, $\Delta=0$, $Q_m=5\times 10^5$ and $\omega_m/2\pi=10^6$ Hz. The shaded regimes
are where entanglement vanishes. \label{ENuni}}
\end{figure}
If we take into account the environmental temperature as shown in Fig. \ref{ENucT} in the introduction part, the unconditional
entanglement vanishes when the temperature is higher than $10$ K given the following specifications:  $\gamma/\omega_m=1$,  $\Delta/\omega_m=-1$, $\Omega_q/\omega_m=1$ and $Q_m=5\times 10^5$ with $\omega_m/2\pi=10^6$ Hz. In contrast, the conditional one exists even when the temperature becomes higher than $100$ K. Therefore, only when the information contained in the cavity output is properly treated will the observer be able to recover the quantum correlation between the oscillator and the cavity mode at high temperature. In fact, the temperature is not the dominant figure that determines the existence of quantum entanglement. A recent investigation showed that
in the simple system with an oscillator interacting with a coherent optical field, quantum entanglement always exists between the oscillator and outgoing optical field \cite{uni}. The resulting entanglement strength only
depends on the ratio between the characteristic interaction strength $\Omega_q$ and thermal-noise strength $\Omega_F$, rather than the environmental temperature. We can make some correspondences to the results in Ref. \cite{uni} by assuming a large cavity bandwidth. In such a case, the cavity mode exchanges information with the external outgoing field at a timescale much shorter than the thermal decoherence timescale of the oscillator. In Fig. \ref{ENuni}, we show the resulting $E_{\cal N}$ of the conditional entanglement as a function of cavity bandwidth and environmental temperature with fixed interaction strength. The entanglement can persist at a very high temperature ($10^4$ K shown in this plot!) as long as the cavity bandwidth is large. This, to some extents, recovers the results obtained in Ref. \cite{uni}.

\section{Effects of imperfections and thermal noise\label{sec7}}

To motivate cavity-assisted cooling experiments, we will consider effects of various imperfections that exist in a
real experiment, which include nonunity quantum efficiency of the photodetection, thermal noise and optical loss.
The effects of nonunity quantum efficiency and thermal noise have already been taken into account in the equations
of motion. With an optical loss, some uncorrelated vacuum fields enter the cavity in an unpredictable way. A small
optical loss will not modify the cavity bandwidth significantly but will introduce an additional force noise, which
is [cf. Eq. (\ref{18})]
\be
S_{FF}^{\rm add}(\Omega)=4\,\hbar\, m\,\Omega_q^3\,\gamma_{\epsilon}\,|\chi|^2(\gamma^2+\Omega^2+\Delta^2),
\ee
where $\gamma_{\epsilon}\equiv c\,\epsilon/(4L) $ is the effective bandwidth that induces by an optical loss
of $\epsilon$. For numerical estimations, we will use the following experimentally achievable parameters:
\ba\nonumber
&m=1\,{\rm mg},\;I_0=3\,{\rm mW},\; {\cal F}=3\times 10^4,\;{\omega_m}/{(2\pi)}=10^5\,{\rm Hz},\\
&Q_m=5\times 10^6,\;L=1\,{\rm cm},\;\eta=0.95,\;\epsilon=10\,{\rm ppm},
\ea
where $\cal F$ is the cavity finesse and $Q_m\equiv \omega_m/(\gamma_m)$ is the mechanical quality factor.
This gives a coupling strength of $\Omega_q/\omega_m\approx0.6$ (for $\Delta=-\omega_m$) and a cavity bandwidth
$\gamma/\omega_m= 2.5$. The final results will not change if we increase both mass and power with the same factor,
which essentially gives the same effective interaction strength.

\begin{figure}
\includegraphics[width=0.65\textwidth, bb= -165 0 510 250, clip]{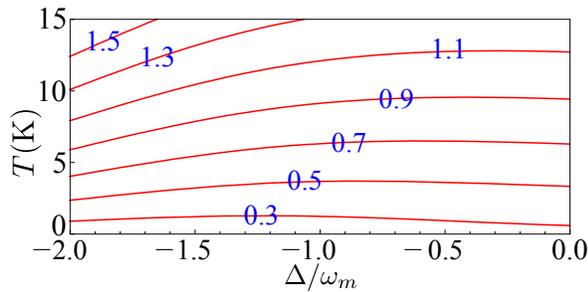}
\caption{Occupation number of the optimally controlled state as a function of the temperature and the
cavity detuning. The other specifications are chosen to be achievable
in a real experiment, which are detailed in the main text. \label{exp}}
\end{figure}
In Fig. \ref{exp}, we show the corresponding occupation number for the controlled state as a function of
environmental temperature and cavity detuning. An occupation number less than one can be achieved when the
environmental temperature becomes lower than 10 K given the above specifications. If the oscillator can
sustain higher optical power, one can increase the interaction strength to reduce thermal excitations.

\section{Conclusion\label{sec8}}
We have shown that both the conditional state and the optimally controlled state of the mechanical oscillator
can achieve a low occupation number even if the cavity bandwidth is large. Therefore, as long as information
of the oscillator motion contained in the cavity output is carefully recovered, the resolved-sideband limit will
not pose a fundamental limit in cavity-assisted cooling experiments.
This work can help understanding the intermediate regime between optomechanical cooling and feedback cooling,
which will be useful for searching optimal parameters for a given experimental setup. In addition,
we have shown that the optomechanical entanglement between the cavity mode and the oscillator can
be significantly enhanced by recovering information, and its existence becomes insensitive to the environmental
temperature.

\section*{Acknowledgements}
We thank T. Corbitt, F. Ya. Khalili, H. Rehbein, and our colleagues at TAPIR and MQM
group for fruitful discussions.  H. M. is supported by the Australian Research Council
and the Department of Education, Science and Training.  S. D., H.M.-E., and Y. C.
are supported by the Alexander von Humboldt Foundation's Sofja Kovalevskaja Programme, NSF grants
PHY-0653653 and PHY-0601459, as well as the David and Barbara Groce startup fund at Caltech.
H. M. would like to thank D. G. Blair, L. Ju and C. Zhao for their keen supportsl of his visit to Caltech.


\section*{References}
\begin{thebibliography}{99}
\bibitem{qmeas} V. B. Braginsky, and F. Ya. Khalili, {\it Quantum Measurement} Publisher: Cambridge Univ. Press (1992).
\bibitem{Bouwmeester} W. Marshall, C. Simon, R. Penrose and D. Bouwmeester, Phys. Rev. Lett. {\bf 90}, 130401 (2003).
\bibitem{Mancini2} S. Mancini, D. Vitali and P. Tombesi, Phys. Rev. Lett. {\bf 90}, 137901 (2003).
\bibitem{Vitali} D. Vitali, S. Gigan, A. Ferreira, H. R. B\"{o}hm, P. Tombesi, A. Guerreiro, V. Vedral, A. Zeilinger
and M. Aspelmeyer Phys. Rev. Lett. {\bf 98}, 030405 (2007).
\bibitem{Paternostro} M. Paternostro, D. Vitali, S. Gigan, M. S. Kim, C. Brukner, J. Eisert and M. Aspelmeyer,
Phys. Rev. Lett. {\bf 99}, 250401 (2007).
\bibitem{Helge} H. M\"{u}ller-Ebhardt, H. Rehbein, R. Schnabel, K. Danzmann, and Y. Chen, Phys. Rev. Lett. {\bf 100}, 013601 (2008).
\bibitem{Hartmann}  M. J. Hartmann and M. B. Plenio, Phys. Rev. Lett. {\bf 101}, 200503 (2008).
\bibitem{Zurek} W. Zurek, Phys. Rev. D {\bf 24}, 1516 (1981); Phys. Rev. D {\bf 26}, 1862 (1982); Phys. Today {\bf 44}, 36 (1991).
\bibitem{Diosi1} L. Di\'{o}si, Phys. Lett. A {\bf 120} 377 (1987); Phys. Rev. A {\bf 40} 1165; J. Phys. A: Math. Theor. {\bf 40} 2989 (2007).
\bibitem{Penrose} R. Penrose, Gen. Rel. Grav. {\bf 28} 581 (1996); Phil. Trans. R. Soc. Lond. A {\bf 356} 1927; {\it The Road to Reality:
A Complete Guide to the Laws of the Universe}, Alfred A.\ Knopf, (2005).
\bibitem{connell} A. D. O'Connell, M. Hofheinz, M. Ansmann, R. C. Bialczak, M. Lenander, E. Lucero,  M. Neeley,
D. Sank, H. Wang, M. Weides, J. Wenner, John M. Martinis, and A. N. Cleland, Nature (London), advance online publication (2010).
\bibitem{Blair} D.  Blair, E. Ivanov, M. Tobar, P. Turner, F. van
Kann, and I. Heng, Phys. Rev. Lett. {\bf 74}, 1908 (1995).
\bibitem{Cohadon} P. Cohadon, A. Heidmann and M. Pinard, Phys. Rev. Lett. {\bf 83}, 3174 (1999).
\bibitem{Metzger}  C. Metzger, and K. Karrai, Nature (London) {\bf 432}, 1002 (2004).
\bibitem{Naik} A. Naik, O. Buu, M. LaHaye, A. D. Armour, A. Clerk, M Blencowe and K. Schwab,
Nature (London) {\bf 443}, 14 (2006).
\bibitem{Gigan} S. Gigan, H. R. B\"{o}hm, M. Paternostro, F. Blaser, G. Langer, J. B. Hertzberg, K. C. Schwab,
D. B\"{a}uerle, M. Aspelmeyer and A. Zeilinger, Nature (London) {\bf 444}, 67 (2006).
\bibitem{Arcizet} O. Arcizet, P. Cohandon, T. Briant, M. Pinard and A. Heidmann, Nature (London) {\bf 444}, 71 (2006).
\bibitem{Kleckner} D. Kleckner and D. Bouwmeester, Nature {\bf 444}, 75 (2006).
\bibitem{Schliesser1} A. Schliesser, P. DelHaye, N. Nooshi, K. Vahala, and
T. Kippenberg, Phys. Rev. Lett. {\bf 97}, 243905 (2006).
\bibitem{Corbitt1}  T. Corbitt, Y. Chen, E. Innerhofer, H. Muller-Ebhardt, D. Ottaway, H. Rehbein, D. Sigg, S. Whitcomb,
C. Wipf and N. Mavalvala, Phys. Rev. Lett. {\bf 98}, 150802 (2007).
\bibitem{Corbitt2} T. Corbitt, C. Wipf, T. Bodiya, D. Ottaway, D. Sigg, N. Smith, S. Whitcomb, and N. Mavalvala,
Phys. Rev. Lett. {\bf 99}, 160801 (2007).
\bibitem{Schliesser2} A. Schliesser, R. Rivire, G. Anetsberger, O. Arcizet, and T. Kippenberg, Nature Physics {\bf 4}, 415 (2008).
\bibitem{Poggio} M. Poggio, C. L. Degen, H. J. Mamin, and D. Rugar, Phys. Rev. Lett. {\bf 99}, 017201 (2007).
\bibitem{Favero} I. Favero, C. Metzger, S. Camerer, D. K\"{o}nig, H. Lorenz, J. Kotthaus and K. Karrai, App. Phys. Let. {\bf 90}, 104101 (2007).
\bibitem{Teufel} J. Teufel, J. Harlow, C. Regal, and K. Lehnert, Phys. Rev. Lett. {\bf 101}, 197203 (2008).
\bibitem{Thompson} J. Thompson, B. Zwickl, A. Jayich, F. Marquardt, S. Girvin and J. Harris, Nature {\bf 452}, 72 (2008).
\bibitem{Lowry} C. Mow-Lowry, A. Mullavey, S. Go$\beta$ler, M. Gray and D. McClelland, Phys. Rev. Lett. {\bf 100}, 010801 (2008).
\bibitem{Groblacher}  S. Gr\"{o}blacher, S. Gigan, H. R. B\"{o}hm, A. Zeilinger, and M. Aspelmeyer, EPL {\bf 81}, 54003 (2008).
\bibitem{Schediwy}  S. W. Schediwy, C. Zhao, L. Ju, D. G. Blair and P. Willems, Phys. Rev. A {\bf 77}, 013813 (2008).
\bibitem{Jourdan} G. Jourdan, F. Comin, and J. Chevrier, Phys. Rev. Lett. {\bf 101}, 133904 (2008)
\bibitem{Aspelmeyer} S. Gr\"{o}blacher, J. Hertzberg, M. Vanner, G. Cole, S. Gigan, K. Schwab and M. Aspelmeyer, Nature Physics {\bf 5}, 485 (2009).
\bibitem{LIGO} B. Abbott {\it et al.} LIGO Scientific Collaboration, New Journal of Physics {\bf 11}, 073032 (2009).
\bibitem{Schwab} T. Rocheleau, T. Ndukum, C. Macklin, J. Hertzberg, A. Clerk and K. Schwab, arXiv:0907.3313v1 (2009).
\bibitem{Vyatchanin_cooling} S. Vyatchanin, Dokl. Akad. Nauk SSSR {\bf 234}, 1295 (1977).
\bibitem{Mancini} S. Mancini, D. Vitali, and P. Tombesi, Phys. Rev. Lett. 80, 688 (1998).
\bibitem{Marquardt} F. Marquardt, J. Chen, A. Clerk, and S. Girvin, Phys. Rev. Lett. {\bf 99}, 093902 (2007).
\bibitem{Rae} I. Wilson-Rae, N. Nooshi, W. Zwerger, and T. Kippenberg, Phys. Rev. Lett. {\bf 99}, 093901 (2007).
\bibitem{Genes} C. Genes, D. Vitali, P. Tombesi, S. Gigan and M. Aspelmeyer, Phys. Rev. A {\bf 77}, 033804 (2008).
\bibitem{ctrl} S. Danilishin, H. M\"{u}ller-Ebhardt, H. Rehbein, K. Somiya, R. Schnabel, K. Danzmann, T. Corbitt,
C. Wipf, N. Mavalvala, and Y. Chen, arXiv:0809.2024 [quant-ph] (2008).
\bibitem{Diosi2} L. Di\'{o}si, Phys. Rev. A {\bf 78}, 021801(R) (2008).
\bibitem{Rabl} P. Rabl, C. Genes, K. Hammerer and M. Aspelmeyer, Phys. Rev. A {\bf 80}, 063819 (2009).
\bibitem{zhao_prl} C. Zhao, L. Ju, H. Miao, S. Gras, Y. Fan and D. Blair, Phys. Rev. Lett. {\bf 102}, 243902 (2009).
\bibitem{QN_int} F. Elste, S. Girvin and A. Clerk, Phys. Rev. Lett. {\bf 102}, 207209 (2009).
\bibitem{Corbitt3} Private communications.
\bibitem{Hopkins} A. Hopkins, K. Jacobs, S. Habib, and K. Schwab, Phys. Rev. B {\bf 68}, 235328 (2003).
\bibitem{Gardiner} C. Gardiner and P. Zoller, {\it Quantum Noise} 3rd ed. Publisher: Springer-Verlag, Berlin (2004).
\bibitem{Milburn} G. Milburn, Quantum Semiclass. Opt. {\bf 8}, 269 (1996).
\bibitem{Doherty1} A. Doherty, S. Tan, A. Parkins, and D. Walls, Phys. Rev. A {\bf 60}, 2380 (1999).
\bibitem{Doherty2} A. Doherty and K. Jacobs, Phys. Rev. A {\bf 60}, 2700 (1999).
\bibitem{state_pre} H. M\"{u}ller-Ebhardt, H. Rehbein, C. Li, Y. Mino, K.
Somiya, R. Schnabel, K. Danzmann and Y. Chen, Phys. Rev. A {\bf 80}, 043802 (2009).
\bibitem{uni} H. Miao, S. Danilishin and Y. Chen, arXiv:0908.1053 [quant-ph] (2009).
\bibitem{q_eraser} M. Scully and K. Dr\"{u}hl, Phys. Rev. A {\bf 25}, 2208 (1982).
\bibitem{q_eraser_exp} Y. Kim, R. Yu, S. Kulik, Y. Shih and M. Scully, Phys. Rev. Lett. {\bf 84}, 1 (2000).
\bibitem{Peres} A. Peres, Phys. Rev. Lett. {\bf 77}, 1413 (1996).
\bibitem{Horodecki} M. Horodecki, P. Horodecki, R. Horodecki, Phys. Lett. A {\bf 223}, 1 (1996).
\bibitem{Simon} R. Simon, Phys. Rev. Lett. {\bf 84}, 2726 (2000).
\bibitem{Vidal} G. Vidal and R. F. Werner, Phys. Rev. A {\bf 65}, 032314 (2002).
\bibitem{Adesso} Review article by G. Adesso and F. Illuminati, J. Phys. A: Math. Theor. {\bf 40} 7821 (2007) and references therein.
\bibitem{scaling_law} A. Buonanno and Y. Chen, Phys. Rev. D {\bf 67}, 062002 (2003).
\bibitem{Kimble} H. J. Kimble, Y. Levin, A. B. Matsko, K. S. Thorne, and S. P. Vyatchanin, Phys. Rev. D {\bf 65}, 022002 (2001) and references therein.
\end{thebibliography}
\end{document}